\documentclass[aps,prb,twocolumn,showpacs,floatfix]{revtex4}
\usepackage{graphicx}% Include figure files
\usepackage{subfigure}
\usepackage{epsfig}
\newcommand{\bs}{\symbol{92}}

\begin{document}

\title{Non-reciprocal optical reflection from a bidimensional
array of \\subwavelength holes in a metallic film}

\author{Micha\"{e}l Sarrazin}
\email{michael.sarrazin@fundp.ac.be} \affiliation{Laboratoire de
Physique du Solide, Facult\'es Universitaires Notre-Dame de la
Paix, \\rue de Bruxelles 61 , B-5000 Namur, Belgium}

\author{Jean Pol Vigneron}
\affiliation{Laboratoire de
Physique du Solide, Facult\'es Universitaires Notre-Dame de la
Paix, \\rue de Bruxelles 61 , B-5000 Namur, Belgium}

\date{\today}

\begin{abstract}
Using simulations and theoretical arguments we investigate the
specular reflection of a perforated gold film deposited on a glass
substrate. A square lattice of cylindrical holes is assumed to
produce the periodic lateral corrugation needed to hybridize the
surface plasmons with radiative states. It is shown that,
contrasting transmission approaches, a knowledge of the reflection
on either side of the film provides separate information on the
gold-vacuum surface plasmons and on the gold-glass interface
plasmons. Recent experimental data on a specific implementation of
this system are reexamined; these show a good agreement between
the measured reflections and the simulations in both directions of
incident wave probes. This confirms the importance of taking into
account the reflection asymmetry in the far-field assessment of
surface plasmons properties.
\end{abstract}

% insert suggested PACS numbers in braces on next line
%% 42.70.Qs -- Photonic bandgap materials
%%%%%%%%%%%%%%%%%%%%%%%%%%%%%%%%%%%%%%%%%%%%%%%%%%

%\maketitle must follow title, authors, abstract, \pacs, and \keywords
\maketitle

In recent years, subwavelength hole arrays in metallic films have
received considerable attention, mainly because of the
experimental findings of Thomas Ebbesen\cite{Ebbesen-Nat-1998}.
Such systems show a particularly intriguing zeroth order optical
transmission : first, the specularly transmitted beam shows an
intensity higher than that of the incident beam filtered by the
geometrical aperture of the holes; second, the transmitted beam
develops a strong characteristic wavelength
dependence\cite{Ebbesen-Nat-1998}. Recent theoretical and
experimental studies have tried to clarify the detailed mechanisms
involved in these phenomena\cite{Ebbesen-Nat-1998,
Ghaemi-PRB-1998,Popov-PRB-2000,Moreno-PRL-2001,Degiron-APL-2002,Sarrazin-PRB-2003,Altewischer-OL-2003,Barnes-PRL-2004}.
In this context, it was rapidly recognized that surface plasmons
(SPs) could be a key to the understanding of the transmission
measurements\cite{Ebbesen-Nat-1998,Ghaemi-PRB-1998,Popov-PRB-2000,Moreno-PRL-2001,
Degiron-APL-2002,Sarrazin-PRB-2003,Altewischer-OL-2003,Barnes-PRL-2004}.

To be more precise, when the incident light illuminates the array
of apertures, many diffraction orders are generated. Some of them
are evanescent and exhibit a resonant coupling with SPs. They
contribute to the zeroth diffraction order as a result of
multiscattering and the strong exaltation of the electromagnetic
field associated with resonant processes results in an enhanced
transmission\cite{Ebbesen-Nat-1998,Ghaemi-PRB-1998,Popov-PRB-2000,Moreno-PRL-2001,
Degiron-APL-2002,Sarrazin-PRB-2003,Altewischer-OL-2003,Barnes-PRL-2004}.
In a recent work\cite{Sarrazin-PRB-2003} we have suggested
redefining the role of SPs in Ebbesen experiments and placing it
in the required context of resonant Wood
anomalies\cite{Wood-PR-1935}. In so doing, we have shown that the
transmission spectrum could be better depicted as a series of Fano
profiles. These recognizable lineshapes result from the
interference of non resonant transfers with resonant transfers
which involve the film eigenmodes and evanescent diffraction
orders. We were then able to draw attention to the fact that each
transmission maximum-minimum pair can be viewed as a Fano
profile\cite{Fano-AnnPhy-1938}. On quantitative grounds, the
interplay of such an assymetric Fano profile actually dissociates
the presence of a maximum or minimum in the transmission spectrum
from the expected location of an eigenmode
resonance\cite{Sarrazin-PRB-2003}.

It is somewhat surprising that reflection has not been as widely
studied as transmission. One of our early contributions to this
field described the theoretical reflection of an array of holes in
a chromium film\cite{Sarrazin-PRB-2003} and, at about the same
time, Altewischer \textit{et al} reported experimental reflection
studies which emphasized its non-reciprocal
properties\cite{Altewischer-OL-2003}. As is well-known, there is a
significant difference between the reflection spectra for light
beams coming from one side or the other side of the device. Such
difference is not observed in the transmission spectra, in
agreement with the reciprocity theorem\cite{Petit80}. The
dissymmetry observed in the reflection spectra usually contains
very significant information and both reflection and transmission
should be thoroughly understood before a full knowledge of the
optical properties of the system can be
claimed\cite{Sarrazin-PRB-2003, Altewischer-OL-2003}. The present
paper proposes a theoretical investigation of the device studied
by Altewischer et al\cite{Altewischer-OL-2003}. This will provide
an opportunity to establish the role of SPs in these experiments
and, at the same time, relate the far-field transfer coefficients
to the very local SPs dynamics.

The simulations in this paper are based on a coupled modes method
which combines a scattering matrix formalism with a plane wave
representation of the fields\cite{Sarrazin-PRB-2003}. The
inherently slow convergence of this representation is here
accelerated by a Li algorithm\cite{Li-JOSAA-1997}. This technique
provides a computation scheme for the amplitude and polarization
($s$ or $p$) of reflected and transmitted fields in any diffracted
order\cite{Sarrazin-PRB-2003}.

Fig.\ref{fig1} describes the system under study, chosen to model
the samples experimentally investigated by
Altewischer\cite{Altewischer-OL-2003}. These samples can be
described as a square lattice grating of circular holes perforated
in a thin gold film deposited on a glass substrate. In accordance
with experiments, $100$ nm is set for the radius of each hole and
the lattice parameter is taken to be $a=700$ nm. The film
thickness is matched to the experimental value of $200$ nm. The
gold permittivity is obtained from experimental
tables\cite{Palik91}. The substrate slab is modelled by a
semi-infinite glass medium with a refractive index of
$\sqrt{\varepsilon _s}=1.51$\cite{Altewischer-OL-2003}. Note that
the experimental device exhibited a very thin bonding layer of
titanium which separates the gold film from the
substrate\cite{Altewischer-OL-2003}. The vanishing thickness of
this film justifies that it will not be considered in our
calculations. The linearly polarized incident light is normal to
the interface. Its electric-field vector is oriented along the
$\left[1,0\right]$ direction, which joins the centers of
nearest-neighboring holes.

\begin{figure}
\caption{A view of the system under
study\cite{Altewischer-OL-2003}. Transmission and reflection are
calculated for zero diffraction order and normal
incidence.}\label{fig1}
\end{figure}

Fig.\ref{fig2}(a) shows the specular reflection as a function of
the incident wavelength in both directions of propagation, i.e. an
incidence from vacuum or from glass. The transmission is also
shown, with a 5-times expansion of the vertical
scale\cite{Altewischer-OL-2003}. The reflection presents strong
variations near wavelengths $757$ nm $(1^{\prime })$, $812$ nm
$(2^{\prime })$ and $1146$ nm $(3^{\prime })$. These variations
are fully consistent with experimental spectral features found at
$747$ nm, $810$ nm and, presumably, at just over $1100$
nm\cite{Altewischer-OL-2003}. Fig.\ref{fig2}(b), \ref{fig2}(c),
and \ref{fig2}(d) reproduces the curves shown on
Fig.\ref{fig2}(a), but exposes them to a comparison with the
corresponding experimental data. Despite the necessary
idealization step in designing the modelled theoretical device, we
note an excellent agreement in all these
comparisons\cite{Altewischer-OL-2003}. In particular, the relative
locations of transmission maxima and reflection minima are
correctly predicted. The only discrepancies which may still raise
questions seem to be the weak oscillations found in all calculated
spectra between $850$ nm and $1050$ nm, not found in experiment.
The spectral line shapes are reminiscent of those described for
similar systems in recent computation\cite{Sarrazin-PRB-2003} and
experimental\cite{Barnes-PRL-2004} reports. The difference between
the reflections on both sides of the film, i.e. the reflection
nonreciprocity, is evident from the comparison of
Fig.\ref{fig2}(c) and  Fig.\ref{fig2}(d). These non-reciprocal
properties of the reflection were also mentioned by Barnes et
al\cite{Barnes-PRL-2004} when describing their optical properties
of structured metallic films.

\begin{figure}
\caption{(color online) (a) : Calculated transmission (curve
marked \textbf{T}) and reflection for zero diffraction order on
the structure of Fig.\ref{fig1}. On reflection curves, the solid
(dashed) line refers to incidence from the vacuum (glass) side.
The reflection nonreciprocity (difference according to incident
directions) takes the form of two peaks at wavelengths $(1^{\prime
})$ and $(2^{\prime })$ for vacuum incidence, while only one at
wavelength $(2^{\prime })$ remains in glass incidence. Other
panels : (b) computed (dashed) transmission and, (c) and (d)
reflections (from vacuum and glass, respectively) compared to
measurements (solid) from Altewischer \textit{et
al}\cite{Altewischer-OL-2003}. Note that experimental values are
not available above $1100$ nm.}\label{fig2}
\end{figure}

In the present investigation the coupling between the various
evanescent p-polarized diffraction orders and the surface plasmons
is of prime importance. Because of the periodic corrugation of the
film, the diffracted surface plasmons are sampled by the
diffracted probe field at many frequencies. The resonant coupling
of surface plasmons to p-polarized diffraction orders at matching
frequencies and surface wave vectors will lead to what will be
called "resonant diffraction orders". For an isolated surface
plasmon which develops on a single interface, the wavelength which
characterizes each resonant diffraction order is given
by\cite{Raether88}
\begin{equation}\label{eq1}
\lambda _{sp}^{u/m\text{ }(i,j)}=\frac a{\sqrt{i^2+j^2}}\text{Re}\sqrt{\frac{%
\varepsilon _u\varepsilon _m}{\varepsilon _u+\varepsilon _m}}.
\end{equation}
In this equation, $\lambda _{sp}={{2\pi c} \mathord{\left/
 {\vphantom {{2\pi c} \omega _{sp} }} \right.
 \kern-\nulldelimiterspace} \omega _{sp} }$ where $\omega _{sp}$ expresses
the resonant angular frequency, $a$ is the grating parameter and
$(i,j)$ denotes the vector ${\bf g}$ of the reciprocal lattice
such that ${\bf g}=(2\pi /a)(i,j)$. $\varepsilon _m$ is the metal
permittivity and $\varepsilon _u$ is either the permittivity of
glass ($\varepsilon _s$) or vacuum ($\varepsilon _v=1$).

For the resonant orders $(\pm 1,0)$ we get at substrate/metal and
vacuum/metal interfaces $\lambda _{sp}^{s/m\text{ }(1,0)} =
1084.6$ nm and $\lambda _{sp}^{v/m\text{ }(1,0)} = 720.5$ nm
respectively. Both wavelengths are indicated by vertical dashes
$(3)$ and $(1)$ respectively in the Fig.\ref{fig2} and
Fig.\ref{fig3}. For the resonant orders $(\pm 1,\pm 1)$ at
substrate/metal interface, we get $\lambda _{sp}^{s/m\text{
}(1,1)}=787.7$ nm indicated by dash $(2)$. Dash $(0)$ corresponds
to the resonant orders $(\pm 2,0)$ at substrate/metal interface
for $\lambda _{sp}^{s/m\text{ }(2,0)} = 608.1$ nm. This resonance
appears in transmission as the first peak at $610$ nm.

The contrast between the reflection spectra for both incidence
directions allows additional information to be obtained on
processes involving specifically the surface plasmons at both
interfaces\cite{Altewischer-OL-2003}. For light arriving from the
vacuum, two dips at the wavelengths $(1^{\prime })$ and
$(2^{\prime })$ are observed. By contrast, for light arriving from
the glass side a significant dip shows up at the wavelength
$(3^{\prime })$ while $(1^{\prime })$ almost disappears.
Altewischer et al\cite{Altewischer-OL-2003} suggested that such an
asymmetry in the reflection spectra could lead to the
identification of the interface on which the SPs are dominantly
excited. We think that, furthermore, these results indicate that
the excited resonances are mainly localized on the interface which
is illuminated first and thus their effects are more easily
perceived with reflection than with transmission.

\begin{figure}
\caption{(color online) Electric field modulus spectrum for
various diffraction orders for light impinging from the vacuum
side (solid line) or from the glass side (dashed line) of the
film. (a) Diffracted field of orders $(\pm 1,0)$ on the
substrate/metal interface. (b) Diffracted field of orders $(\pm
1,0)$ at the vacuum/metal interface. (c) Diffracted field of
orders $(\pm 1,\pm 1)$ at the substrate/metal interface. The
amplitude of the incident field is taken to be $1$ $V.m^{-1}$. (d)
Absorption from the incident wave against wavelength, for
incidences from the vacuum side (solid line) and from the glass
side (dashed line).}\label{fig3}
\end{figure}

To confirm and detail this interpretation, we have computed the
amplitudes of main resonant diffraction orders as a function of
wavelength and illumination directions. The field amplitude is
considered here as a measure of the degree of excitation of a
surface plasmon located on a specific interface. Fig.\ref{fig3}a
shows the local amplitude of resonant orders $(\pm 1,0)$ at the
substrate/metal interface. When the light crosses the
substrate/metal interface at the first place, it shows a peak
amplitude at the wavelength $(3)$, predicted by Eq.\ref{eq1}. One
notes a second peak at wavelength $(3^{\prime })$ close to $(3)$,
possibly related to interferences of a true resonance with a
non-resonant process, as found in Fano
resonances\cite{Sarrazin-PRB-2003}. These peaks can easily be
associated with the dip at similar wavelengths in the reflection
spectrum. By contrast, when the vacuum/metal interface is
illuminated first, the amplitude of these resonances collapses
with the result that there is an attenuation of the associated dip
in the reflection spectrum.

Fig.\ref{fig3}b shows resonant orders $(\pm 1,0) $ on the
vacuum/metal interface. Here the resonance develops at the
wavelength $(1)$ and we note the existence of two other peaks at
wavelengths $(1^{\prime })$ and $(2^{\prime })$. It is likely that
these peaks arise from a surface plasmon splitting due to the
coupling of the $(\pm 1,0)$ plasmon orders at the vacuum/metal
interface with the $(\pm 1,\pm 1)$ plasmon orders at the
substrate/metal interface. If the vacuum/metal interface is
illuminated first, amplitudes of peaks $(1)$ and $(1^{\prime })$
are stronger than in the opposite illumination. This explains the
nonreciprocity found in the reflection spectra for wavelengths
close to $(1^{\prime })$. The spectral region close to the
wavelength $(2^{\prime })$, of the resonant order amplitude, does
not significantly change when changing the propagation direction,
leading to a weak nonreciprocity effect in the reflection
spectrum. In Fig.3c we show the amplitude of orders $(\pm 1,\pm
1)$ at the substrate/metal interface and focus on the resonance
wavelength noted $(2)$. A major contrast is obtained, with a large
field amplitude under the direct illumination of the
substrate/metal interface and a much weaker amplitude under
illumination from the vacuum side.

The amplitude of the resonances of orders $(\pm 1,\pm 1)$ at the
substrate/metal interface and $(\pm 1,0)$ at the vacuum/metal
interface are influencing each other because of the proximity of
their resonance wavelengths. Indeed, the profile of the orders
$(\pm 1,0)$ amplitude at vacuum/metal interface, just after the
wavelength $(1)$, can be understood as a Fano profile. Such a Fano
resonance can be interpreted as resulting from the interference of
the orders $(\pm 1,0)$ resonance at the vacuum/metal interface,
and the orders $(\pm 1,\pm 1)$ resonance at the substrate/metal
interface which then appears as a result of multiscattering of
$(\pm 1,0)$ into $(\pm 1,\pm 1)$. This explains, for instance, why
the wavelengths $(1^{\prime })$ and $(2^{\prime })$ do not
strictly correspond to the resonance wavelength of the different
resonant orders. Fano profiles resulting from the interference of
each order as a result of multiscattering lead to a visible
resonance wavelength shift\cite{Sarrazin-PRB-2003}.

We note that this analysis of the comparison between surface
plasmon features and reflection corroborate the interpretation of
Altewischer et al\cite{Altewischer-OL-2003}, from which useful
information about surface plasmons can be drawn from the study of
the reflection nonreciprocity.

In addition to transmission and reflection, it might be useful to
investigate the loss response function for both directions of
illumination. This is shown in Fig.\ref{fig3}d. As light comes
from the vacuum side, we observe two sharp peaks at wavelength
$(1^{\prime })$ and $(2^{\prime })$. These peaks correspond to the
surface plasmon resonance already described above. On the other
hand, when the light enters through the glass side, the first peak
collapses, whereas another large double peak appears at the
wavelength $(3^{\prime })$, in perfect correlation with the
surface plasmon excitation. This, again, is in full agreement with
the experimental results of Barnes et al\cite{Barnes-PRL-2004} and
previous computations\cite{Sarrazin-PRB-2003}, confirming that
both loss peaks correspond to a surface plasmon resonance. The
loss line shapes can easily be correlated with the dominant
features of the reflection spectra. These features indicate that,
when SPs are excited, a large part of their energy is dissipated
via the Joule effect, leading to lower reflection intensities. In
addition, a smaller part of SPs energy is transferred to the
transmitted beams as a result of multiscattering. So, as SPs are
essentially excited on the interface which is encountered first,
the reflection spectrum is dominated by processes associated with
the losses from the plasmons carried by this interface.

As already noted by Barnes\cite{Barnes-PRL-2004}, this is quite
different for transmission, where SPs of both interfaces equally
contribute to enhanced transmission whichever side is illuminated
first. The interface which receives the incident beam provides
strong SPs resonances. Such resonances only give a weak
contribution to the transmission since the associated field is
damped when propagating through the holes before providing
transmitted intensity. On the opposite interface, SPs are excited
by an incident light weakened by its propagation through the
holes, so that SPs on the opposite interface are only weakly
excited. Their contribution to the transmitted field is thus also
weak. In other words, SPs of the illuminated interface are
strongly excited and then damped after propagation through the
holes, whereas SPs on the opposite interface are weakly excited as
the incident light is first damped. The result is that SPs of both
interfaces contribute to the transmission with similar magnitudes
and that the choice of the direction of propagation does not
influence the information content of the measured transmission.

In conclusion, the examination of several experimental
studies\cite{Altewischer-OL-2003,Barnes-PRL-2004} in the light of
the interpretation scheme presented
elsewhere\cite{Sarrazin-PRB-2003} reinforces the point of view
expressed in this scheme by exhibiting qualitative and
quantitative agreements between measured and computed light
filtering spectra of a perforated gold layer covering a glass
substrate. A particular effort was made to study the
nonreciprocity property of the reflection. It was found that the
asymmetry relative to the light propagation direction is
effectively caused by the localization of SPs resonances at gold
film surfaces and that this can be exploited to investigate
separately the gold/glass and gold/vacuum plasmon coupling to
radiative waves. The transfer functions computed in detail in this
paper agree well with the quoted experimental data, available to
us at the time of writing. One of the outcomes of these
experiments is the role played by the corrugated-surface plasmons
in the build up of the light transport process. In order to shed
some light on this issue, the present work also looked into the
theoretical value of the excited field at the location of the
metal/vacuum and metal/glass interfaces and determined the degree
of excitation of the interface plasmons. The strong correlation
found between the plasmon amplitude and the reflection variations
shows unambiguously the role played by the metal surface
collective excitations in the light transfer process. This
conclusion suggests that the experiments could be taken some steps
further by considering complementary measurements, at least on the
outer vacuum/gold surface, of the far field in reflectance
experiments and the near field with the scanning near-field
optical microscope. A correlation between these separate
investigations should be highly instrumental in elucidating the
mechanisms of the light transfer processes.

%%%%%%%%%%%%%%%%%%%%%%%%%%%%%%%%%%%%%%%%%%%%%%%%%%%%
\begin{acknowledgments}

We thank M.P. van Exter and E. Altewischer for enlightening advice
and for making extensive experimental data available to us during
the preparation of this work and P. Kelly for his advice on the manuscript.

We acknowledge the use of the Namur Interuniversity Scientific
Computing Facility (Namur-ISCF), a joint project between the
Belgian National Fund for Scientific Research (FNRS), and the
Facult\'{e}s Universitaires Notre-Dame de la Paix (FUNDP).

This work was partially supported by the EU Belgian-French
INTERREG III project \bs PREMIO'', the EU5 Centre of Excellence
ICAI-CT-2000-70029 and the Inter-University Attraction Pole (IUAP
P5/1) on \bs Quantum-size effects in nanostructured materials'' of
the Belgian Office for Scientific, Technical, and Cultural
Affairs.

\end{acknowledgments}

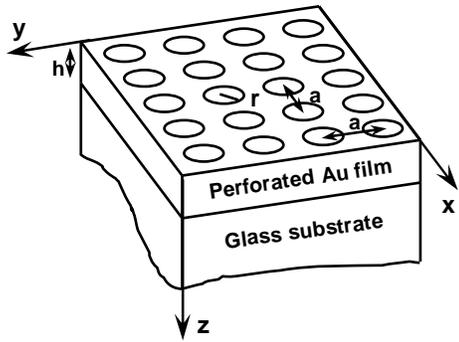
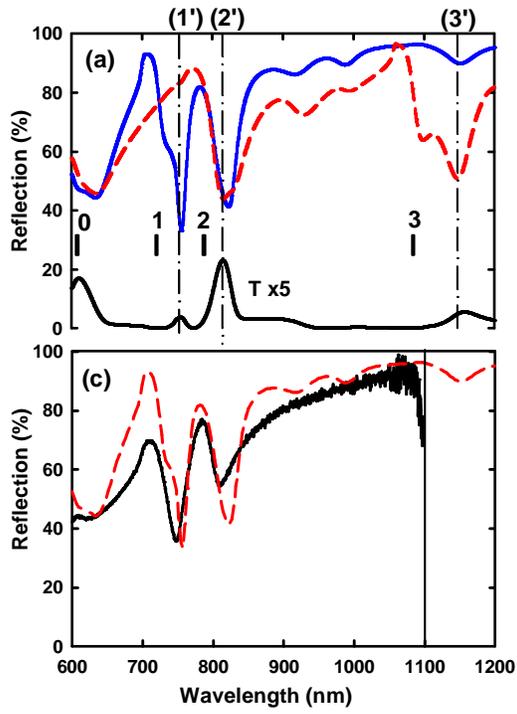
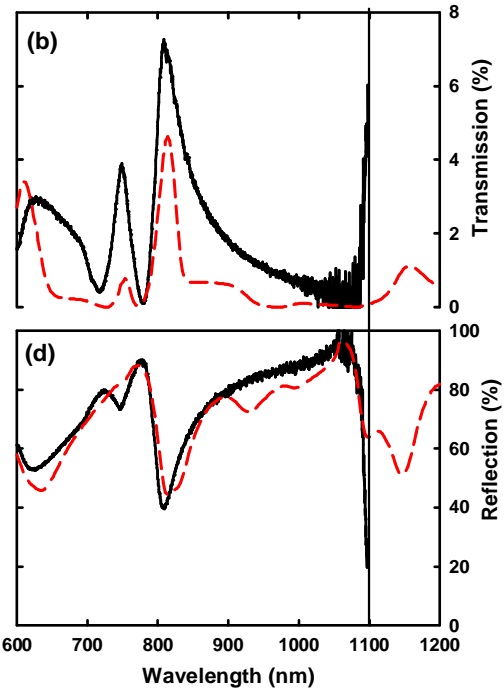
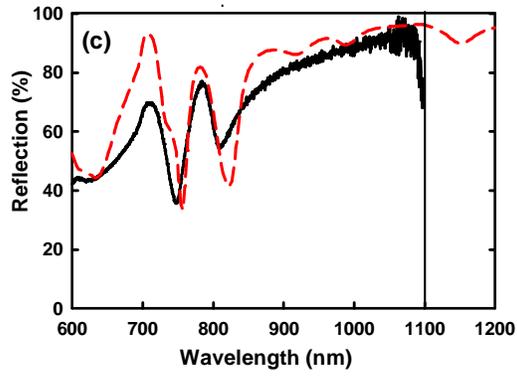
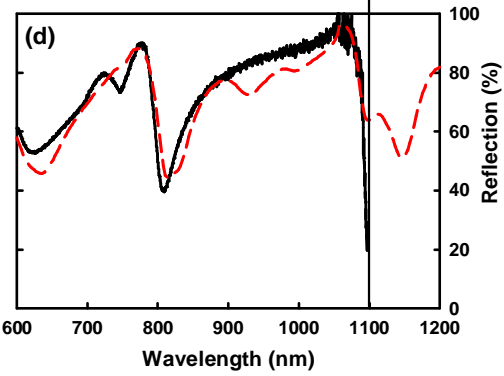
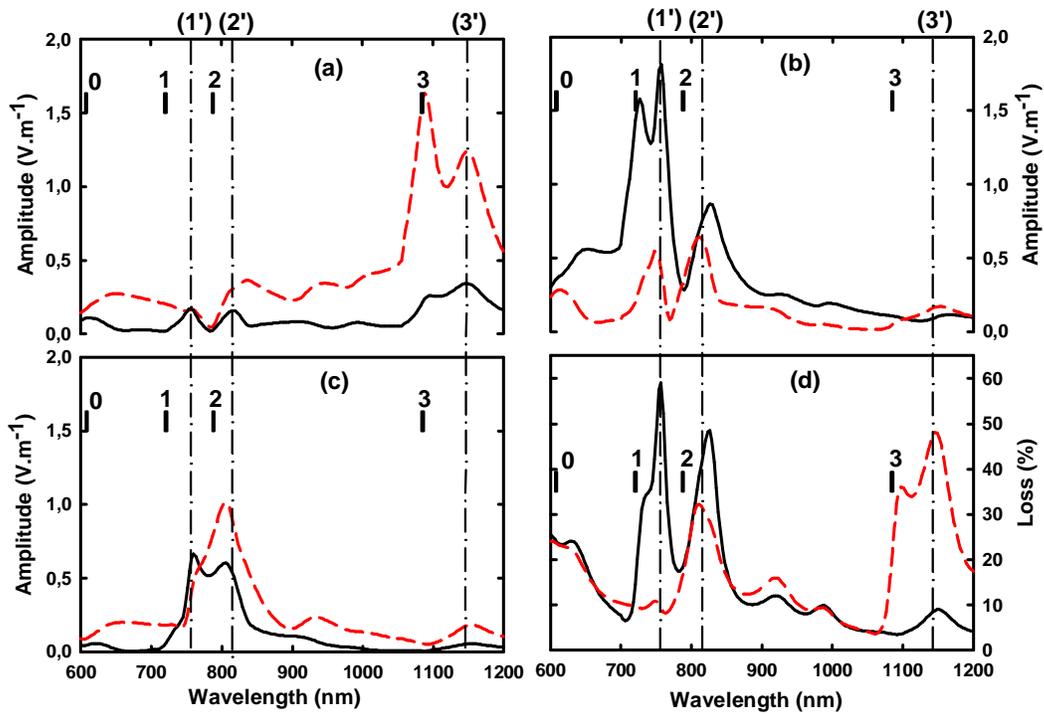
\end{document}